\documentclass[aps,prb,superscriptaddress,twocolumn]{revtex4-1}
\usepackage{amsmath}
\usepackage{color}
\usepackage{floatrow}
\usepackage{amssymb}
\usepackage{graphicx}
\usepackage{xcolor}
\usepackage{blindtext}

\def\vr{{\bf r}}
	
	\def\EE{{\bf E}}

\begin{document}

\title  {Simple correction to bandgap in IV and III-V semiconductors:  \\ an improved first-principles local density functional theory}
\author{Sujoy Datta}
\affiliation{Department of Physics, University of Calcutta, Kolkata 700009, India}
\affiliation{Department of Physics, Lady Brabourne College, Kolkata 700017, India}
\author{Prashant Singh}
\email{prashant40179@gmail.com}
\affiliation{Ames Laboratory, U.S. Department of Energy, Iowa State University, Ames, Iowa 50011 USA}
\author{Chhanda B. Chaudhuri}
\affiliation{Department of Physics, Lady Brabourne College, Kolkata 700017, India}
\author{Debnarayan Jana}
\affiliation{Department of Physics, University of Calcutta, Kolkata 700009, India}
\author{Manoj K. Harbola}
\affiliation{Department of Physics, Indian Institute of Technology, Kanpur, 208016, India}
\author{Duane D. Johnson} \email{ddj@iastate.edu; ddj@ameslab.gov}
\affiliation{Ames Laboratory, U.S. Department of Energy, Iowa State University, Ames, Iowa 50011 USA}
\affiliation{Department of Materials Science $\&$ Engineering, Iowa State University, Ames, Iowa 50011, USA}
\author{Abhijit Mookerjee}
\affiliation{Department of Physics, Lady Brabourne College, Kolkata 700017, India}
\affiliation{S. N. Bose National Centre for Basic Sciences, Salt Lake City, Kolkata 700098, India}

\begin{abstract}
We report results from a fast, efficient, and first-principles full-potential  N$^{th}$-order muffin-tin orbital (FP-NMTO) method combined with van Leeuwen-Baerends correction to local density exchange-correlation potential. We show that more complete and compact basis set is critical in improving the electronic and structural properties. We exemplify the self-consistent FP-NMTO calculations on group IV and III-V semiconductors. Notably, predicted bandgaps, lattice constants, and bulk moduli are in good agreement with experiments (e.g., we find for Ge  $0.86~e$V, $5.57$~\AA, $75$~GPa  vs. measured $0.74~e$V, $5.66$~\AA, $77.2$~GPa). We also showcase its application to the electronic properties of 2-dimensional $h-$BN and $h-$SiC, again finding good agreement with experiments.
\end{abstract}

\date{\today}
\maketitle

\section{Introduction}
Semiconducting materials remain of great interest due to their central role in modern electronics.\cite{Droge2000,Landemark1994,Stroscio1986,Feenstra1987,Ku2002,Smith2012,Yutaka2009} In this era of designing optimal materials, it has become essential to estimate quickly and accurately  their bandgaps. Experimentally we would benefit from an efficient computational tool for predicting bandgaps (among many other properties) prior to synthesis. We consider the problem of predicting bandgaps to be ``solved'' by a method that delivers an accurate result for compounds spanning the whole periodic table and is simultaneously practical to compute. Standard exchange correlation (XC) functionals to density functional theory (DFT), such as local density (LDA) or semi-local generalized gradient (GGA) \cite{GGA2,GGA4} approximations, have been the backbone of DFT calculations for decades. Unfortunately, the unphysical self-Coulomb repulsion\cite{Perdew1981} leads to a systematic underestimation of bandgaps.\cite{Perdew1983,Sham1983,Mori2008} 

{\par}To characterize the electronic properties of a semiconductor or insulator, the fundamental bandgap ($E_{gap}$) is a key quantity.  $E_{gap}$ is defined as the difference of ionization energy ($I$) and electron affinity ($A$) of the $N$ electron system. While the Kohn-Sham (KS) gap ($E_{gap}^{KS}$) of non-interacting ($N$) electron system is defined as the difference of the highest occupied (HO=$-I$) and the lowest unoccupied (LU=$-A$) band energies. $E_{gap}^{KS}$ would be same as fundamental gap if  $I$ and $A$ are exact; however, KS-DFT using LDA or GGA leads to a large underestimation of $E_{gap}^{KS}$.\cite{GGA2,GGA4,Perdew2017, Seidl1996} Thus, the fundamental gap is defined as a discontinuous change ($\Delta_{xc}$) in the KS potential: $E_{gap} = E_{gap}^{KS} + \Delta_{xc}$.\cite{Godby1988} The $\Delta_{xc}$ in DFT is refer to as the jump discontinuity in the derivative of the total energy with respect to $N$.\cite{Perdew1982,MKH1998} Although the LDA exchange energy has a derivative discontinuity at integer $N$, the potential shows no such discontinuity.\cite{AKK2008} For LDA or GGA,\cite{GGA2,GGA4} the discontinuity can 50\% or larger and no straightforward method exists to estimate $\Delta_{xc}$.\cite{Singh2013} Due to the $\Delta_{xc}$ contribution to $E_{gap}$ in KS-DFT, there is pressing need for ``improved" functionals to assess quickly the HOMO-LUMO from KS orbitals that is closer to  observed bandgaps. To access to the ``exact" KS-DFT functional (i.e., not LDA/GGA), the proposed functional should reproduce the  $\Delta_{xc}$ in the KS potential.\cite{Trushin2016} 

{\par}Several improved semi-local XC functionals have been constructed using increasingly complex density dependence, but satisfy some exact physical features.\cite{Medvedev2017,Yu2016} If an approximate (semi-local) functional can reproduce the exact asymptotic behavior in potential, then it is good enough for extracting excited state information.\cite{Savin, Becke2006, vLB} Some semi-empirical functionals are derived as correction to  LDA exchange,\cite{mbj,pb2018,mbjp,mbj1,mbj2,AKK2008,Armiento2013,Singh2013,Singh2015,Singh2016,Becke2006} incorporating density gradients to yield asymptotically well-behaved potentials,\cite{Singh2013,Singh2015,Singh2016,BS2017,Singh2017} which are easily handled mathematically and yield good results with little cost computationally.\cite{mbj,Singh2016,Singh2017} Yet, care is needed in getting discontinuities in the solid-state limit.\cite{JCP2015} Similarly, accurate, variational exact-exchange functionals approaching the van Leeuwen-Baerends (vLB) corrected LDA form have appeared,\cite{Chachiyo2017} but are unexplored for solids. 

{\par}Here we use the  vLB-corrected LDA  (LDA+vLB)  in full-potential N$^{th}$-order muffin-tin orbital (FP-NMTO) method. The self-consistent FP-NMTO uses compact basis set to describe the entire system including the interstitial, and it is accurate whether or not it is close-packed. The error arising when making the energy-independent basis set is minimized in FP-NMTO, which is in contrast to Tight-Binding Linearized MTO (TB-LMTO-ASA).\cite{Basisfunction,Basisfunction1,LMTO,LMTO1,Nohara} The FP-NMTO utilizes the interpolation of higher-order energy derivatives of Taylor's expansion of basis set\cite{EMTO,NMTO1,LMTO3,Lambrecht} and provides a smaller and energy-independent basis set. The FP-NMTO basis set is comparable to localized Wannier functions -- a prerequisite for computational speed and efficiency, which we exemplify for group IV and III-V semiconductors. 

\section{Computational Method}
{\par}We apply self-consistent FP-NMTO within LDA+vLB to calculate electronic properties of group IV and III-V semiconductors, and discuss Ge and GaAs. We compare FP-NMTO bandgaps using LDA,\cite{VWN}  PBE,\cite{GGA2} and LDA+vLB\cite{vLB} with experiments and other existing theories. In FP-NMTO, we use third-order correction to the energy derivatives of wave functions ($N=3$) to make the energy-independent basis set. We use $8\times8\times8$ k-point mesh (29 irreducible points) for Brillouin zone sampling. The self-consistent potential converged to a difference of 10$^{-6}$ after several tens of iterations. For lattice constants, we utilized a least square fit of our data to Murnaghan's equation of state.\cite{Murnaghan} The FP-NMTO LDA+vLB equilibrium lattice constants are in good agreement with the  experiments.\cite{alat1,alat2}  All  calculations are done self-consistently and non-relativistically, using Andersen mixing scheme to facilitate convergence. The tetrahedron method was used for k-space integration. (See appendix for more description of TB-LMTO-ASA.)

\section{Theoretical Details}\label{Sec:Theory}
\subsection*{Partial-wave solutions: NMTO vs LMTO}
\begin{figure}[t]
\includegraphics[scale=0.18]{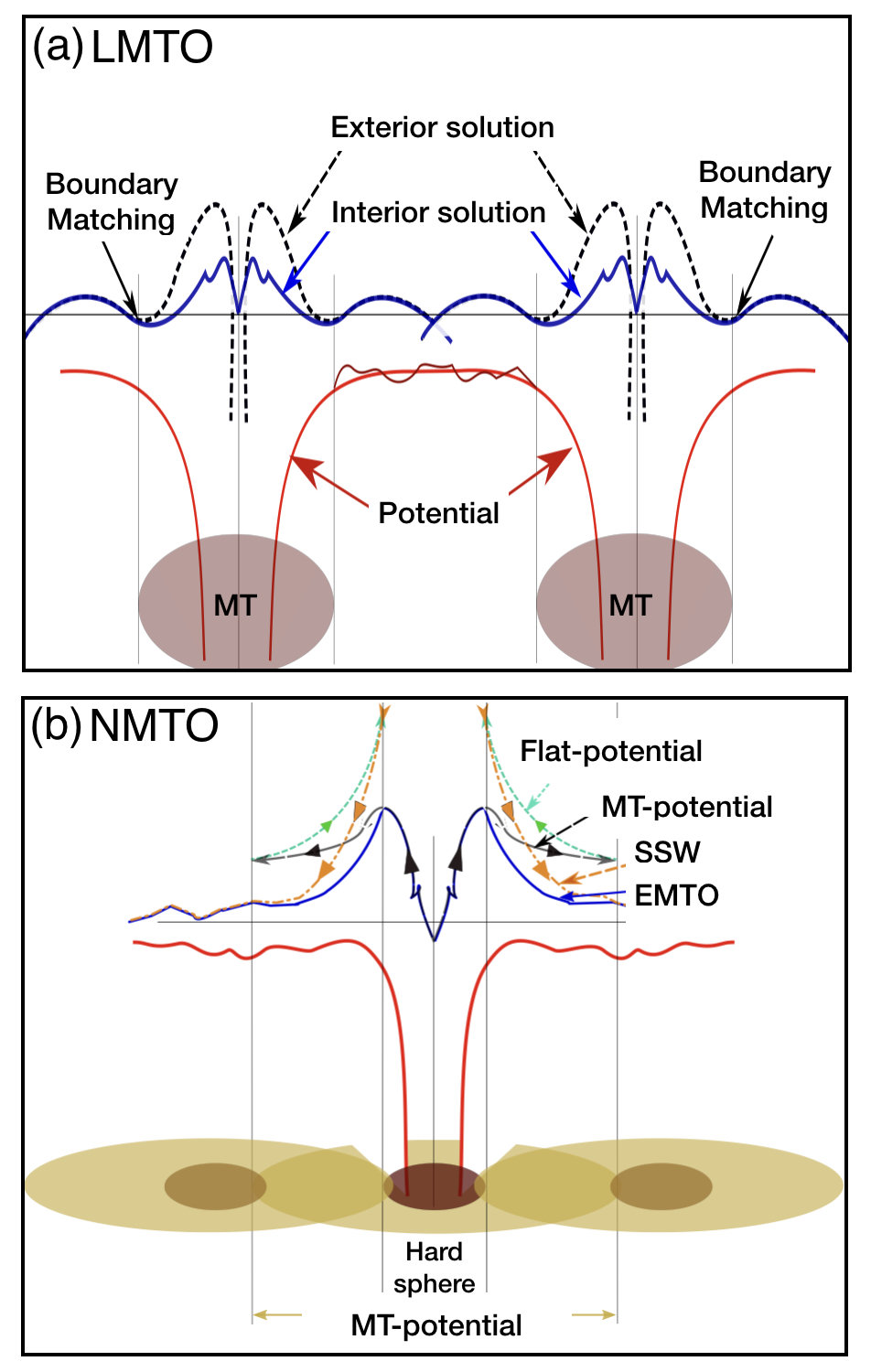}
\caption{(Color online) Schematic of atomic spheres and potential in (a) TB-LMTO-ASA and (b) FP-NMTO.}
\label{fig_lmnm}
\end{figure}

{\par}In the TB-LMTO-ASA, atomic spheres are used to describe both the variation of electronic charge and the potential. These spheres need to be space filling with  minimal (optimal) overlap ($\le15\%$). Overlap is essential for atomic bonding, whereas too large an overlap of charge may lead to its over-counting. The potential for a given atom is reasonably long ranged but the atomic charge density is more localized around the ion cores. However, NMTO has two different classes of concentric spheres: potential (MT/ASA) spheres of radius $s$ and charge spheres (screening/hard spheres) of radius $a$, whereas $a < s$, see Fig.\ref{fig_lmnm}(b). However, the overlap of potential spheres in NMTO can be as large as $50\%$ so full potential approach is a pre-requisite both in atomic spheres and interstitials.

{\par}The solutions of radial KS equations [Eq.~\ref{kseq}] for atomic-potential $\varphi^{as}$ is continued from the core of atom towards the muffin-tin surfaces at $r=s$. This is matched with the solution for a flat part of potential (as in LMTO, this flat potential is taken as zero potential) $\varphi^{int}$, which is denoted by $\varphi^0$. The $\varphi^0$ is valid up to the boundary of hard sphere surface ($r=a$). We normalize at the boundary of hard sphere surface ($a$) such that $\varphi^0|_a=1$. The normalized wave-functions have superscript ``a'':
\begin{eqnarray}
\varphi_{R,l}^a (\epsilon,r) \equiv \frac{\varphi_{R,l}(\epsilon,r)}{\varphi_{R,l}^0(\epsilon,a_R)} \nonumber \\
\varphi_{R,l}^{0 a}(\epsilon,r) \equiv \frac{\varphi_{R,l}^0(\epsilon,r)}{\varphi_{R,l}^0(\epsilon,a_R)} \label{phinorm}
\end{eqnarray}
The $\varphi_{Rl}^{0 a}(\epsilon,r)$ is matched with screened spherical waves (SSW: $\psi^a$) continuously but with a kink at $a$ and  truncated in the region $a_R \leq r \leq s_R$. $\varphi_{Rl}(\epsilon,r)$ is truncated outside $0 \leq r \leq s_R$. The SSWs grow outwards, which are sorted in two groups depending on the values of (R,l,m): active channels (subscript $A \equiv RL$) and passive channels (subscript $I \equiv RL$). The matching described above is for the active channels which are truncated inside hard spheres with kinks. The passive channels are substituted smoothly by $\varphi^{as}$ inside hard spheres. Finally, a basis set with elements termed as Kinked Partial Waves (KPW), which are defined in all space, is formed as:
\begin{eqnarray}
\phi^{a}_{Rlm}(\epsilon, r_R) & = [\varphi_{R,l}^a(\epsilon,r)_R  -  \varphi_{R,l}^{0 a}(\epsilon,r_R)]Y_{l,m}(\hat{r}_R)  \nonumber\\
& +\psi_{Rlm}^a(\epsilon, r_R)
\end{eqnarray}
A linear combination  $\sum_{RL \in A} \phi_{Rlm}^a(\epsilon_i, r_R)c^a_{Rlm,i}$ of KPWs is a solution of KS equation but not the individual KPWs. A kink cancellation condition (similar to the KKR Green's function zero-determinant condition) for solution can be achieved such that a kink in any hard sphere is cancelled by the sum of kinks from tails of KPWs from neighboring sites.

{\par}This basis set spanned by the KPWs is explicitly energy dependent. Energy-dependent basis set slows down the calculation, which was the problem in exact MTO (EMTO) formalism. In LMTO, energy dependence is removed by a self-consistent choice of energy $\epsilon_{\nu}$ about which a Tailor series expansion of basis functions are performed. The series is truncated after first-order term leading to an error $\propto(\epsilon-\epsilon_{\nu})^2$ in the solution of KS equation. The energy-independent set of basis functions, the so-called NMTO are the superposition of KPWs at a mesh of energies (calculated self-consistently) using Newton's interpolation method such that:
 \begin{eqnarray}
  |\chi^{0...N}\rangle &=& \sum_{n=0}^N|\phi(\epsilon_n)\rangle L^{0...N}_n \nonumber\\
  & = & |\phi[0]\rangle+|\phi[01]\rangle(E^{(0...N)}-\epsilon_0)  + \ldots +  \nonumber\\
    & & |\phi[0...N]\rangle(E^{(N-1,N)}-\epsilon_{N-1})(E^{(0...N)}-\epsilon_0) \nonumber
  \end{eqnarray}
where $\phi[0]\equiv\phi(\epsilon_0); \phi[01]\equiv \frac{\phi(\epsilon_0)-\phi(\epsilon_1)}{\epsilon_0-\epsilon_1}$ are the terms in divided difference table. 
The error in the solution of KS equation becomes
\begin{equation}
\psi(\vr)-\psi(\epsilon,\vr) \propto(\epsilon-\epsilon_0)(\epsilon-\epsilon_1)...(\epsilon-\epsilon_n)  .
\end{equation}

\subsection*{van Leeuwen-Baerends Correction to Exchange}
From the Hohenberg-Kohn theorem, the Schr\"{o}dinger's equation for many-body system reduces to the KS equation for a system of non-interacting electrons moving in an effective potential due to all other electrons and ions: 
\begin{eqnarray}
{\bf T}[\rho(\vr)] + {\bf V}[(\vr)]\ \Psi(\vr) = {\bf \mathcal{E}}\ \Psi(\vr).
\label{kseq}
\end{eqnarray}
The total energy-functional having the contribution of this effective potentials is given by
\begin{eqnarray}
\EE  & = &   -{\sum_{\lambda\in occ}\ {\frac{1}{2}}\int d\vr\  {\Psi_\lambda}^* (\vr)\nabla^2 \Psi_\lambda(\vr) } + \int d\vr \rho(\vr) V_{ie}(\vr)  \nonumber\\
  &+&  \frac{1}{2} \iint d\vr d\vr^\prime \frac{\rho(\vr) \rho(\vr^\prime)} {\mid{\vr}-{\vr}^{\prime}\mid}   +  E_{xc}[\rho(\vr)]  
  \label{TotalE}
\end{eqnarray}
\noindent where 1$^{st}$ term is kinetic energy. The 2$^{nd}$, 3$^{rd}$, and 4$^{th}$ terms involve the external ion-electron [$ie$], the Hartree, and exchange-correlation [XC] contributions, respectively, which combine to provide the self-consistently determined effective potential, V$_{eff} ({\bf r})$, written as
\begin{eqnarray}
V_{\rm eff}({\bf r})&\ =\ & V_H[\rho({\bf r})] +V_{ie}[\rho({\bf r})] + V_{xc}[\rho({\bf r})]     .
\label{eq5}
\end{eqnarray}
\noindent
Generally, $V_{xc}[\rho({\bf r})]$ is given by the variation of the $E_{xc}[\rho(\vr)]$ with respect to density $\rho(\vr)$. The  LDA, first introduced by Slater, is based on homogeneous electron gas.\cite{LDA} It simplifies the non-local exchange energy in terms of the local $\rho ^{1/3}$ potential,\cite{Sla51} as  $E_{xc}^{LDA}\simeq \int {\rho ({\vr})\varepsilon _{xc}}\left[ {\rho ({\vr})}\right] {d{\vr}}$. In semi-local form, it is simplified as $E_{xc}^{semi}\simeq \int {\rho ({\vr})\varepsilon _{xc}}[{\rho({\vr})}; {\nabla_{}^{(n)}} \rho({\vr})] {d{\vr}}$  in terms of local XC energy density.
 While calculations using DFT potentials from LDA/GGA are the most successful and widespread in the last three decades, most of these calculations have resulted in significant underestimation of the bandgaps of semiconductors and insulators.

{\par}Many well-established DFT approximations fail to reproduce the observed energy gaps of semiconducting material due to wrong asymptotic  behavior at r$\rightarrow{0}$ and r$\rightarrow{\infty}$ limits.\cite{Singh2016} The LDA potential decays exponentially similar to density for finite systems, however, the potential should decay as $-1/r$, i.e., Coulomb like.\cite{Almb} Becke proposed a nonlocal correction to the energy-functional and then found the exchange potential, \cite{Becke2006} which gave a correct asymptotic form of exchange-energy density but failed to reproduce the exact behavior for the potential. Perdew and Wang developed GGA functionals.\cite{GGA, GGA4} Later, van Leeuwen and Baerends provides a similar type of correction to LDA exchange, significantly improving the eigenvalues for the highest-occupied orbitals of atoms.\cite{vLB,MKH1989,MKH1990,AB1999} 

{\par} In short, the  vLB-corrected LDA potential can be written as
\begin{eqnarray}
v_{xc}(\vr)= [v_{x}^{LDA}(\vr)+ v_{x}^{vLB}(\vr)]+v_{c}^{LDA}(\vr) ,  
\label{eq6}
\end{eqnarray}
where, using $z=\frac{|{\nabla} \rho(\vr) |} {\rho ^{4/3}(\vr)}$ and $\beta = 0.05$,
\begin{eqnarray}
v_{x}^{LB}(\vr)= -\beta \rho^{1/3}(\vr) \dfrac{z^2}{1+3 \beta\ z \sinh^{-1} (z)}
\label{eq7}
\end{eqnarray}
This method of gradient correction to the XC-functional is more built into the theory and thus less artificial than self-interaction-corrected (SIC) approaches.\cite{Singh2016,Singh2017} 

{\par}Singh, et al. \cite{Singh2016,Singh2017} implemented and discussed results for LDA+vLB within LMTO-ASA, in particular, the importance of the vLB-matching condition in the interstitial (the ``asymptotic'' value within a solid).  That is,  different electronic-structure (e.g., site-centered basis vs. plane-wave basis) methods have different reference zeroes for the potentials, i.e., $v_o$. Notably, here, any site-centered method can choose its potential zero inside the crystal, typically in the interstitial to define the stationary wave. As such, we set the LDA+vLB potential to $v_o$ and  enforce its ``asymptotic'' ($-1/r$) behavior in the interstitial. Hence, all potentials (i.e., Eq.~\ref{eq5}) used to solve the KS equations and eigenvalues (dispersion) are defined relative to $v_o$. And, notably, we only require differences of $I$ and $A$ in reference to  $v_o$, which cancels out.   For approximate densities/potentials, a variational form of $v_o$ is useful, as it approaches (to second-order in density error) the KS kinetic energy (dispersion) of exact full-potential results.\cite{PRB.85.2012} As noted in Ref.~\onlinecite{Singh2016}, and as discussed in our Results Section, changing the reference to atomic zero, say, to compare to plane-wave results, one requires the work (dielectric) function $W \propto \epsilon$. Hence, an advantage of the site-centered basis is that it avoids the calculation of $\epsilon$ for some quantities like bandgap, but recovers the same result. In any case, LDA+vLB can be implemented using, e.g.,  {\tt LibXC} software\cite {LibXC2012}, if the asymptotic behavior of  $V_{\rm eff}({\bf r}\rightarrow s)$ is set judiciously in the interstitial of the solid.

\subsection*{   Potential-Shape Effects } 
{\par}The search for a fast, semilocal, multiplicative XC-potential (which is more universally accurate than those presented before) is certainly not an easy task. However, it may be helpful to understand what is going on in terms of the shape of the potentials considered in this work. The vLB correction is motivated by Becke,\cite{BEK88} which also leads to correct $-1/r$ behavior in the outer regions of finite systems. The potential has been employed in the past to study the effect of the correct asymptotic behavior of the potential on response properties of atoms.\cite{vLB} This shows that an improved XC potential that is particularly accurate in the outer regions of the AS gives significantly improved results over the LDA. 

\begin{figure}[t]
\includegraphics[scale=0.15]{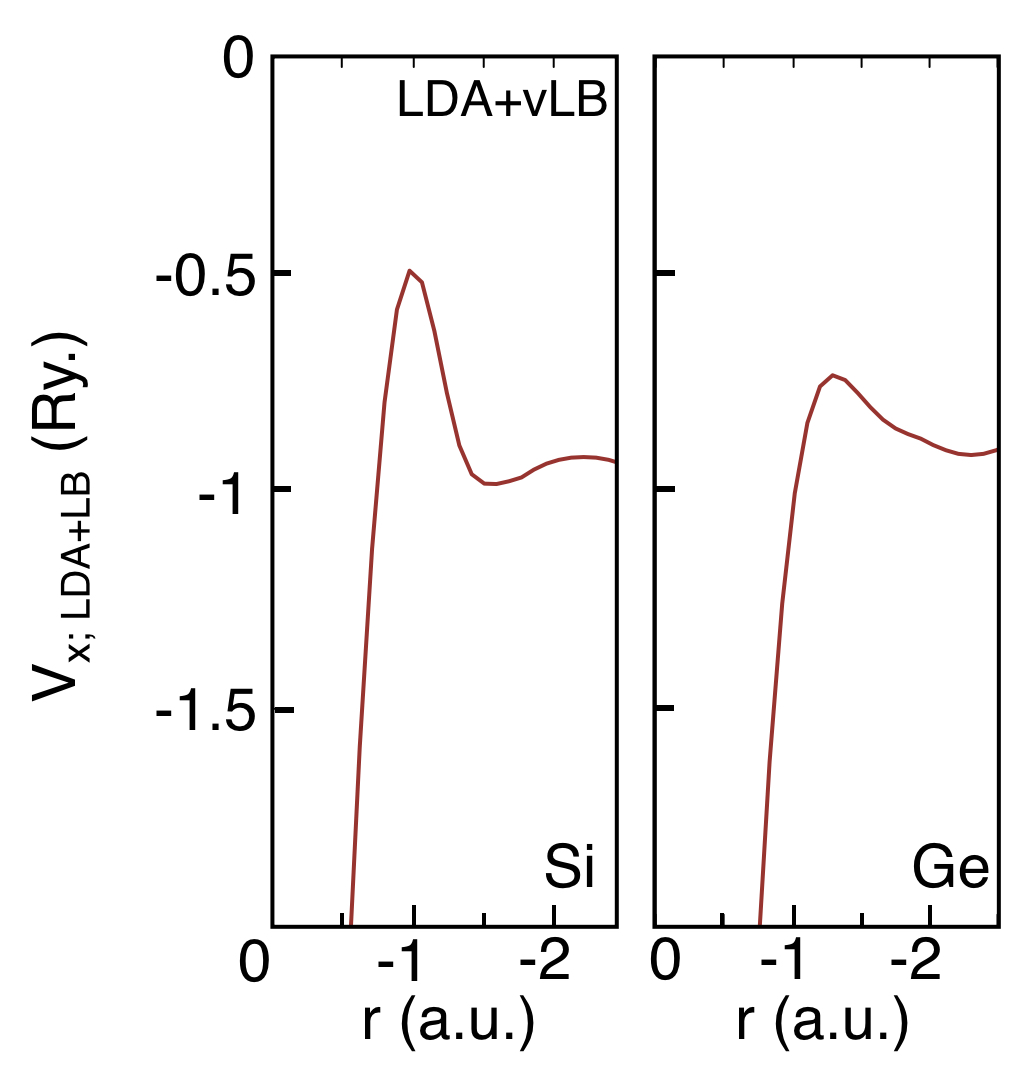} 
\caption{(Color online) vLB-exchange $v_{x}^{LB}(\vr)$ in FP-NMTO for Si and Ge showing good agreement with exact-exchange.\cite{Schilfgaarde2006}}
\label{Ge_Si}
\end{figure}

{\par}The correctness of the eigenvalue of the highest occupied state, the appropriate asymptotic behavior and integer discontinuity of the exchange-correlation potential are all inter-related, which helps to estimate the accurate bandgaps in DFT methods. It is well known that Harbola-Sahani exchange-only functional gives  accurate upper most eigenvalues and correct asymptotic behavior for the finite systems too.\cite{Singh2013,Singh2016,Singh2012} Similar to the inbuilt derivative discontinuity of exact-exchange\cite{Schilfgaarde2006} and Harbola-Sahani exchange-only potential,\cite{Singh2013}  the vLB also shows similar discontinuities at the ``shell-steps" in potential (Fig~\ref{Ge_Si}). Thus, it is not a coincidence that for a large number of systems the exact-exchange, self-interaction correction, and Harbola-Sahni exchange-only potential give significantly improved bandgaps compared to semi-local functionals.\cite{Schilfgaarde2006,Singh2013}  For Ge and Si, we plot (Fig.~\ref{Ge_Si}) the potential in the atomic sphere. The result provides some guidance when choosing (within the KS method) an exchange-correlation potential that is adequate for the problem at hand, e.g., bandgap calculation in our case. Trends in the results could be understood by comparing the shape of the potentials. 

{\par}The discontinuous shift in the exchange potential turns out to be closely related to the ``step structure'' of the exchange potential of atoms, also shown by van Leeuwen-Baerends in the proposed model potential.\cite{vLB} Such a step is present regardless of how small the occupancy of a shell is, as long as it is greater than zero. Hence, in the fractionally-filled orbital case, the orbitals are filled with a successively increasing fractional particle number and a new step is created at the exact point when a new shell is opened. If a boundary condition v$_{X}\rightarrow{0}$  as $r \rightarrow\infty$ is enforced for the potential, it shifts the whole potential discontinuously. Krieger, Li, and Iafrate \cite{KLI1992} observed that with just a tiny fraction of occupancy in a new shell, the exchange potential is shifted discontinuously, where an exact-exchange potential (e.g., van Leeuwen-Baerends) shows the required potential discontinuity.\cite{KLI1992,KLI1990,KK2008} Similar considerations also apply for the time-dependent extension of DFT.\cite{LK2005,MK2005}

\section{Results and discussion}\label{Sec:Res}
{\par}We demonstrate that improving the LDA for its behavior in the outer regions of the atomic sphere and basis set indeed leads to significant improvement in bandgap. We exemplify this by calculating bandgaps of Si, Ge, GaP, GaSb, GaAS, InAs, InSb and InP and compared them with existing theory and experiments. For better understanding, we discuss Ge and GaAs results in  detail.

\begin{figure}[b]
\includegraphics[scale=0.15]{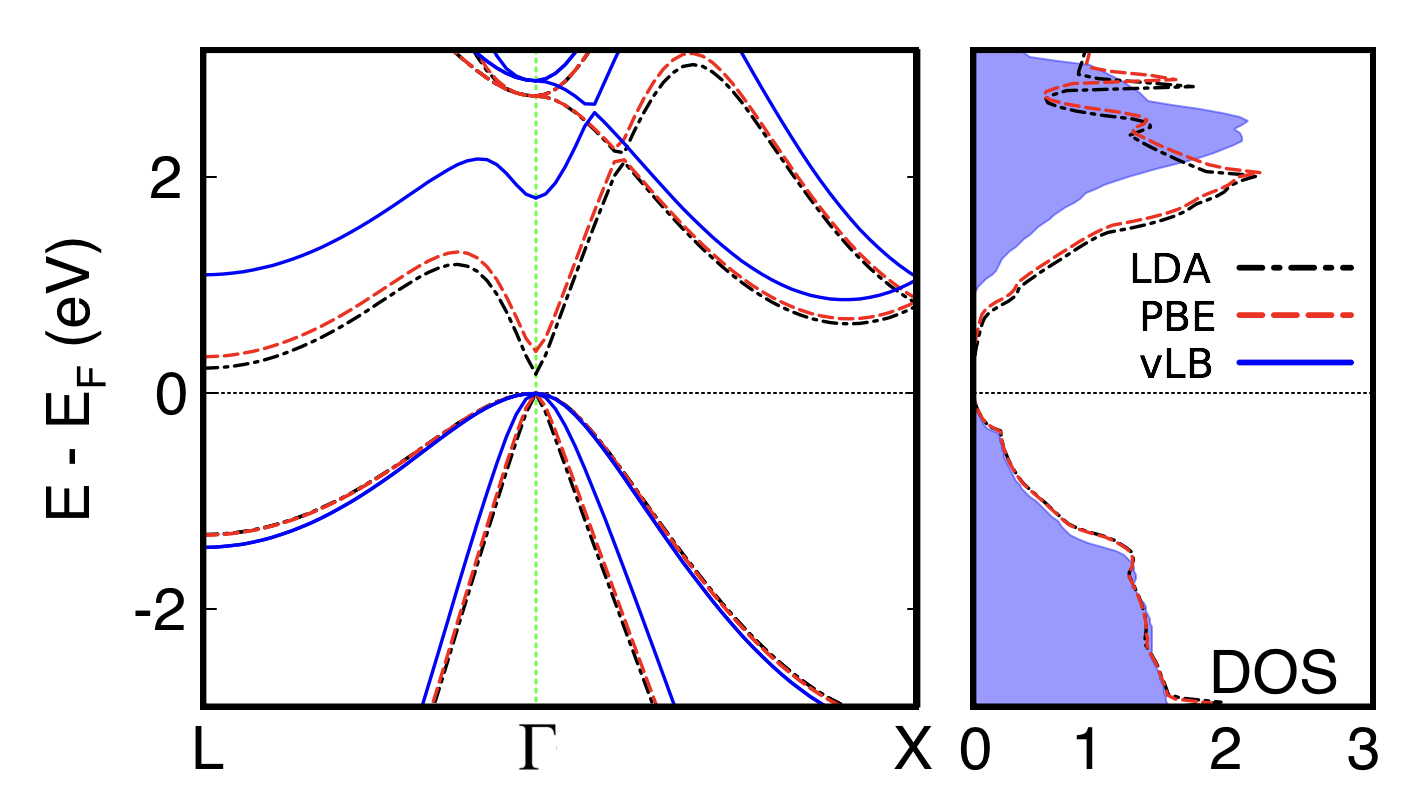} \\ 
 \includegraphics[scale=0.13]{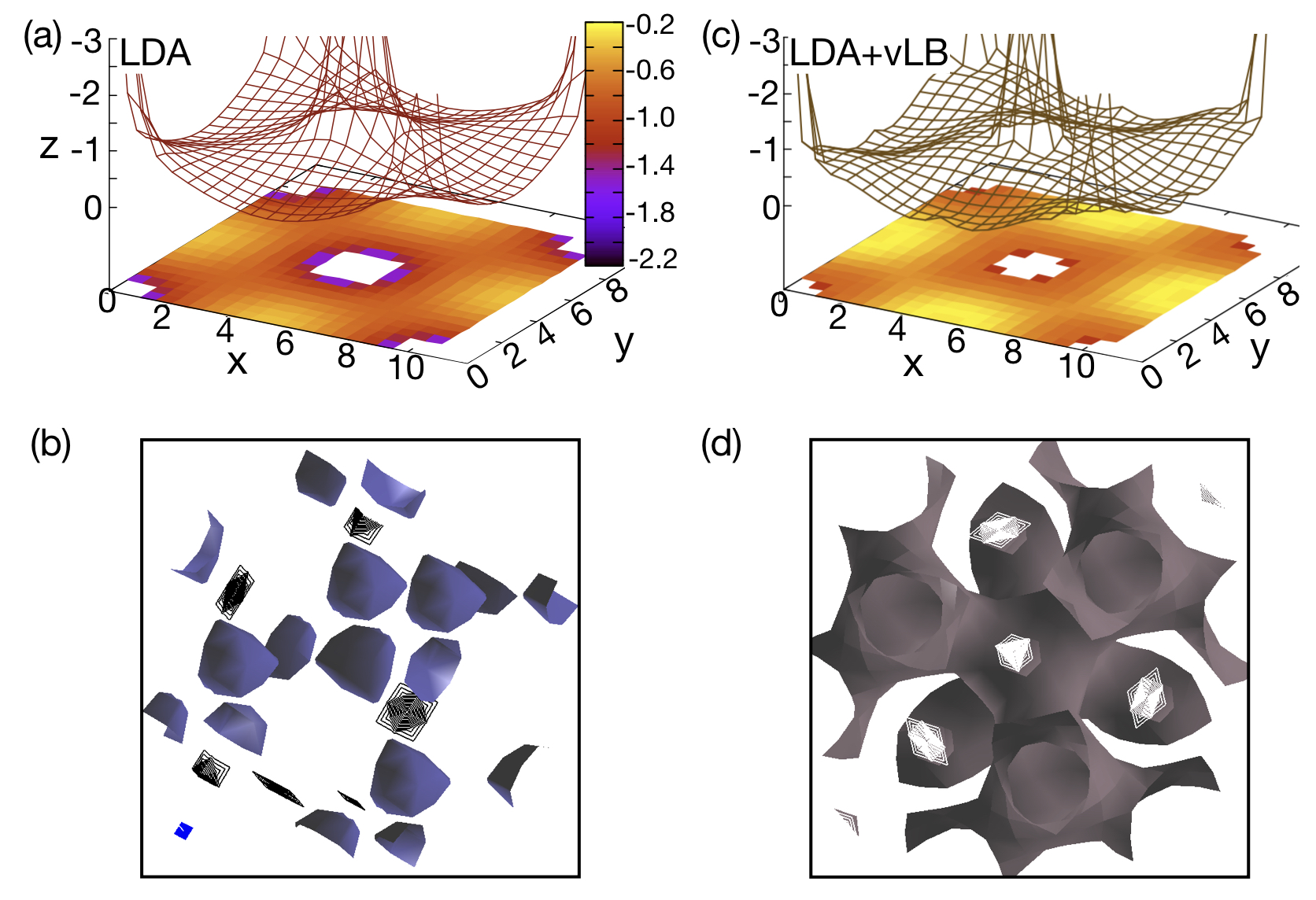}
\caption{(Color online) For Ge, (top) Self-consistent FP-NMTO band structure (L$-\Gamma-$X) and density of states is shown from LDA, PBE, and LDA+vLB. (bottom) FP-NMTO XC potential (z-axis is range) for (a) LDA and (c) LDA+vLB in $\left[1 0 0\right]$-plane. Potential iso-surface plots ($-0.45 e$V/C)  using  (b) LDA and (d) LDA+vLB: XC potential is less local for LDA+vLB, which improves band energies and gap.}
\label{Ge}
\end{figure}

{\it Bulk Germanium}:~In Fig.~\ref{Ge}, we plot the band structure and total density of states (DOS) for Ge. Ge crystallizes in the diamond ($Fd{\bar{3}}m$) structure.\cite{Ge_} Two equivalent Ge's are placed at $(0,0,0)$ and $(1/4,1/4,1/4)$) in a cubic cell with empty spheres (ES) at $(-1/4,-1/4,-1/4)$ and $(1/2,1/2,1/2)$. Both TB-LMTO-ASA and FP-NMTO use average Wigner-Seitz radius 1.387~\AA~ for Ge. For FP-NMTO, the hard sphere radius is 0.971~\AA. The FP-NMTO-LDA gives a direct bandgap of $0.19$ eV at $\Gamma$, which shows small improvement over LMTO+LDA of $0.10$ eV. With the interstitial region treated in the same way, use of the improved FP-NMTO basis set gets better values. However, if we switch to FP-NMTO-PBE, the bandgap is improved to $0.33$ eV, but the nature of bandgap remains direct, and more than 50\% underestimated with respect to the experiments ($0.74$~eV). In contrast, the FP-NMTO using LDA+vLB  increases the bandgap to $0.86$ eV. The key point is that the correct nature of bandgap, which is indirect along $\Gamma\rightarrow{L}$, now matches experiment. The valence bands remain almost same, whereas the conduction bands shift away from E$_{F}$.

{\par} For Ge in LDA and GGA, see Fig.~\ref{Ge}, shows no key difference in band structure. The spaghetti of bands originated from the Ge-4$s$ level were seen between $-13$ eV and $-8.7$ eV. The Ge-4$p$ levels with a contribution of the Ge-4$s$ level toward the less binding energies are found at the region closer to the Fermi energy (E$_{F}$). These bands extend from $-8.7$ eV up to 0 eV. The conduction band mainly results from Ge-4$p$ with minor contributions of Ge-4$s$ atomic orbitals  toward the least excited states of Ge-4$d$ from approximately $6.30$ eV and up. Some degree of degeneracy is also observed depending on which direction of the BZ is traversed: in the  $L \rightarrow \Gamma$, and $\Gamma\rightarrow {X}$ direction. These degenerate bands belong to Ge and filled with  Ge-$p$ electrons. The $\Gamma$ point has three degenerate bands; and the X point at the edge of the BZ  shows two values of energies with a twofold degeneracy.

{\par}Our results show that the fundamental gap of Ge is an indirect one with the maximum of the valence band occurring at $\Gamma$ and the minimum of the conduction band  at the L point.\cite{Landemark1994} The calculated indirect gap at equilibrium lattice constant is 0.86 eV (versus 0.74 eV in experiments). The calculated direct gap at the $\Gamma$ point is $\sim$2.0 eV. The LDA and GGA lead to the similar direct gaps of 0.19 eV and 0.33 eV, respectively. The LDA+vLB structural properties and bandgaps are in better agreement with experiments. This also shows improvement over beyond-DFT approaches.\cite{Ku2002,Smith2012,Rohlfing1993,Hybertsen1986} 

{\par} As the vLB potential was not derived as a functional derivative,  the corresponding energy functional is not yet known, as is so with other functionals. However, the exchange energy can be evaluated by applying the virial theorem based Levy-Perdew sum rule, and with this the prescription for performing DFT calculations for structural minimization for a vLB potential is complete.\cite{LP1985, BH1999}

\begin{figure}[b]
\includegraphics[scale=0.15]{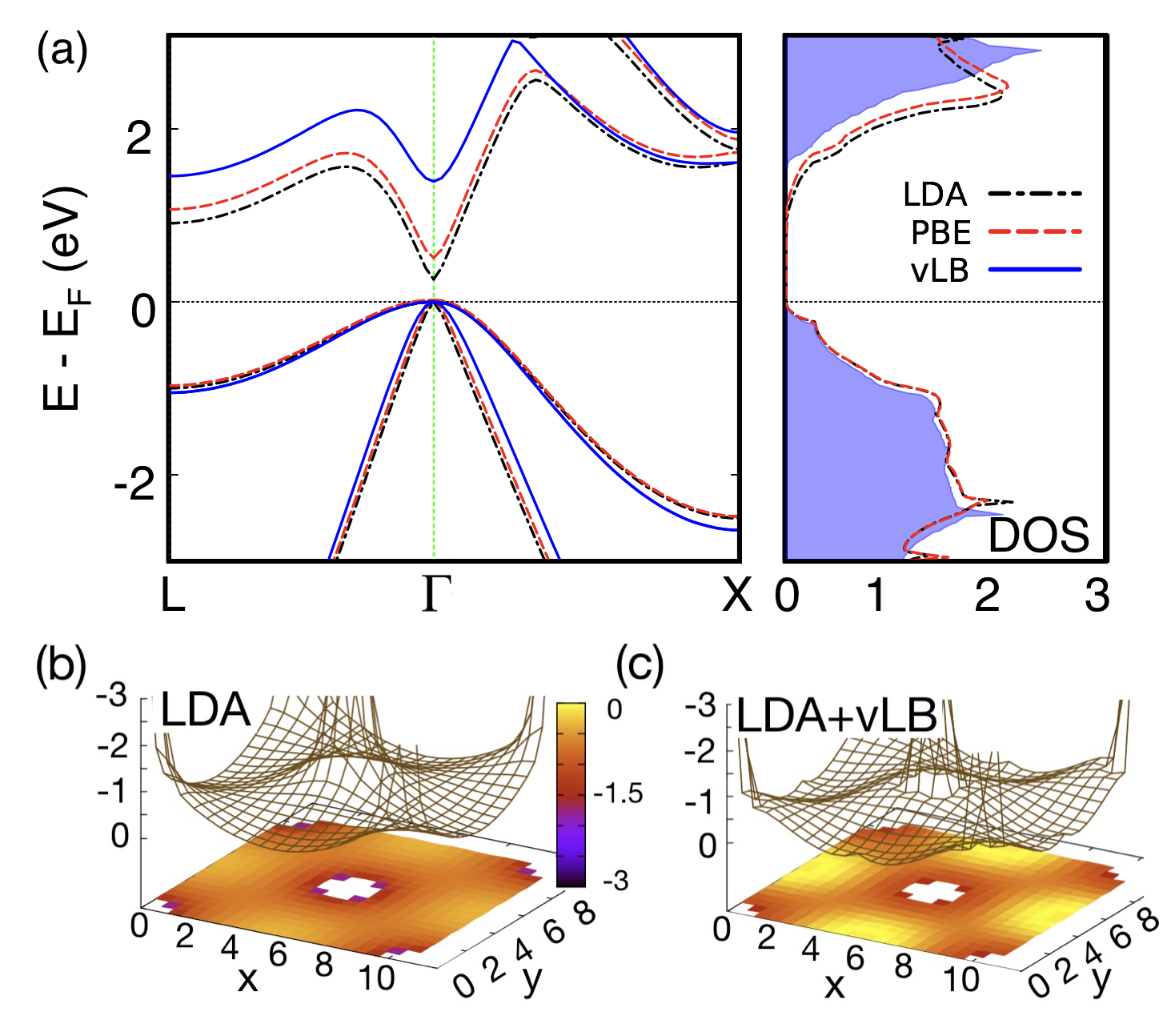}
\caption{(Color online) For GaAs, FP-NMTO  (a) band structure along L$-\Gamma-$X (left-panel) and density of states (right-panel) with LDA, PBE and LDA+vLB. The FP-NMTO-vLB  bandgap (1.43 eV) shows agreement with experiment (1.52 eV).\cite{GaAs2012} We also show the XC-potential surfaces corresponding to (b) LDA and (c) LDA+vLB in $\left[1 0 0\right]$ plane.}
\label{GaAs}
\end{figure}

\begin{table*}[t]
\begin{tabular}{ccccccccccccccc}
\hline
  &System                                                     & Si           & Ge        & GaP     & GaSb   & GaAs            & InAs         & InSb   & InP  \\  \hline
 &a(\AA) ($\rightarrow$)                                              & 5.50 & 5.57 & 5.45 & 6.00 & 5.65 & 6.04 & 6.48 & 5.85 \\ \hline
      XC($\downarrow$) &                                            &  &  &  Band Gap (eV)  &  &  &  &  &  \\ \hline
LDA	(LMTO)			                         & &  0.49(I)  & 0.10  & 1.67  & 0.52 & 0.08 & 0.01  & 0.06  & 0.6 \\
vLB	(LMTO)			                         & &  1.21(I)  & 0.06  & 1.46  & 0.05 & 0.04 &  0.01 &  0.01 & 0.5  \\   \hline
LDA	(NMTO)						& &  0.79(I)  &  0.19(D) &1.53   & 0.25     & 0.33            & 0.01            & 0.00     & 0.5 \\ 
PBE	(NMTO)						& & 0.85(I)   & 0.33(D)  & 1.68  & 0.51     & 0.54            & 0.04            & 0.02     & 0.90  \\ 
vLB	(NMTO)						& & 1.25(I)    & 0.86(I)   & 1.87  & 0.94     & 1.43             & 0.39            & 0.79     & 1.18 \\ \hline
\bf{Experiment}						& &1.17(I)     & 0.74(I)   & 2.32  & 0.81     & 1.52             & 0.43            & 0.23     & 1.42 \\ \hline
mBJ\cite{mbj,pb2018,mbjp,mbj2,mbj1} 	& & 1.15  	   & 0.83      & 2.25   & 0.95  & {\scriptsize 1.53-1.64} & {\scriptsize 0.42-0.67} & {\scriptsize 0.25-0.47} & {\scriptsize 1.42-1.62}  \\ 
LB94\cite{pb2018}     			        & & 0.25       & 0.00       & 0.61  & 0.00 & 0.00 & 0.00 & 0.00 & 0.00 \\ 
GW\cite{Rohlfing1993,2,d,8}			&	           & {\scriptsize 1.29-1.31} & {\scriptsize 0.65-0.71} & 2.80 & 0.62 & 1.58 & 0.31 & 0.08 & 1.44 \\ 
HGH\cite{wang} 					&                  & --	 & --     & 2.44 & 0.55 & 1.01 & 0.19 & 0.21 & 1.23  \\
GDFT\cite{remediakis} 				& &1.17  	  & 0.49  	& 2.47 & 0.93 & 1.72 & 1.40 & 0.99 & 2.55    \\ 
HSE\cite{HSE,HSE1} 				& & 1.28 	  &  0.00	& 2.47 & 0.72 & 1.21 & 0.39 & 0.29 & 1.64 \\
TPSS\cite{tao} 						& & 0.82 	  & 0.56    & 1.98 & 0.08 & 0.52 & 0.00 & 0.00 & 0.90  \\ 
HISS\cite{hiss} 						& & 1.22   	  & 0.54    & 2.67 & 1.31 & 1.86 & 0.93 & 0.80 & 2.23     \\ 
EXX\cite{exx}						& & 1.50   	  & 1.01    & -- 	   & --     & 1.82 & -- 	   & --      & --    \\ \hline
\end{tabular}
\caption{FP-NMTO bandgaps for group IV and III-V materials (in eV) from LDA, PBE and LDA+vLB compared with other theories and experiments. For LMTO-vLB see ref.[~\citenum{Singh2016}]. I=indirect-bandgap; D=direct-bandgap.}\label{tab1}
\end{table*}

{\par}The FP-NMTO  calculated DOS, compared in  Fig.~\ref{Ge} (right-panel), for vLB-corrected shows differences both in the energy positions and in the valence bandwidths with respect to LDA and PBE. We can see a weak shoulder at $-$0.60 eV followed by relatively broader peaks at $-$1.35, $-$2.25, and $-$2.9 eV and a sharper peak at $-$4.7 eV. In the conduction bands, a small shoulder can be found at 0.9 eV followed by a peak at 2.50 eV, with other peaks  at 3.2, 4.0, and 4.8 eV. These DOS peak positions  are in closer agreement with experiments.\cite{Landemark1994,Eastman1975,Straub1985,Muzychenko2010,Jackson1986} Although Ge is known to be an $sp$ material, a correct treatment of 3$d$ states is critical for accurate results. We found that major contribution of Ge-3$d$ states goes to conduction bands with smaller contribution to valence states.  We consider Ge-1$s$2$s$2$p$ into the core, \cite{Rohlfing1993,Shirley1992} which leads to a better description of the band energies. Clearly, improving the basis set by including Ge-3$d$ states into valence achieves better energy description and (indirect) gap of Ge.

{\par}To further elaborate why vLB-corrected LDA in  FP-NMTO produces better result, we plot XC-potential surface for LDA and LDA+vLB of Ge $\left[100\right]$ plane in bottom panel of Fig~\ref{Ge}. The $z-$axis represents the strength of XC-potential in the range $\{0, -3\}$. As the nature of interstitial region can not be understood for the full range of XC-potential,  we show a truncated plot. At atomic sites, the XC-potential forms deep wells, and falls exponentially to $0$ in LDA (Fig.\ref{Ge}(a)). However, the LDA+vLB shows much slower fall than LDA (Fig.\ref{Ge}(c)). The curvature of the XC-potential surface also improves accordingly. The plots are done at isosurface values of $-0.45~e$V/C for both LDA and LDA+vLB in Fig.\ref{Ge}(b),(d). This shows that the XC-potential for Ge in LDA is much more localized at the atoms, but LDA+vLB is more spread over the cell.

{\par}{\it Bulk Gallium-arsenide}:~GaAs crystallizes in the zinc-blende ($F{\bar{4}}3m$) structure,\cite{,GaAs2012} where Ga and As-atoms occupy $(0,0,0)$ and $(1/4,1/4,1/4)$ sites, respectively. The ES are placed to maintain the curvature of potential. We plot FP-NMTO calculated band structure and DOS in Fig.~\ref{GaAs}(a).The bandgap from LDA, PBE and LDA+vLB comes out to be 0.33 eV, 0.54 eV, and 1.43 eV, respectively. The FP-LMTO+vLB calculated bandgap of 1.43 eV shows good agreement with experimentally observed bandgap of 1.52 eV.\cite{GaAs2012} Clearly, LDA and PBE hugely underestimate the bandgap compared to LDA+vLB. Moreover, the direct (D) nature of the bandgap is also well reproduced. The band structure and DOS in Fig.~\ref{GaAs}does not show any constant shift, which rules out any scissor operator behavior of LDA+vLB. The band structure shows larger improvement in energy levels at $\Gamma$ than at X or L points. We see no change in shape of DOS except at the conduction bands near E$_{F}$ and valence bands at $-2.25$ eV. In Fig.~\ref{GaAs} (b),(c), the potential-energy surface for GaAs is shown along $\left[100\right]$, where overlap between the potential spheres are determined self consistently. The $z-$axis shows the strength of XC potential. For clarity, we show a truncated XC-potential surface in the range $\{0, -3\}$ along z$-$axis. We can see that at atomic sites, the XC potential forms deep well like structure, and falls exponentially to $0$, which is similar to Ge. The light yellow zone in Fig.~\ref{GaAs}(c), shows smaller interstitial contribution. This is only possible if we have smaller number of interstitial electrons than LDA (darker spots), which consequently helps correcting the energy levels in conduction band.

{\par}We summarize the bandgaps of IV and III-V semiconductors in Table~\ref{tab1}. The results show that improving the basis set (localized, compact and complete) in conjunction with LDA+vLB yields significantly better results, especially in comparison to modified Becke-Johnson (mBJ), GW, and hybrid functionals. 
So, a better localized basis with no approximations (Linear vs N$^{th}$-order MTO) is important, but the improve description of exchange and it asymptotic behavior is equally necessary (LDA vs LDA+vLB).

Comparing the various results in Table~\ref{tab1}, it is evident that the improvement in basis set affects significantly the results --  for example, compare Si, Ge, and GaSb between LDA and LDA+vLB and LMTO-ASA vs FP-NMTO. (Keep in mind that LMTO-ASA has always used ES to improve the spherical basis for semiconductor materials, which is why it does alright.)  In all cases, the improved basis set is critical; although, when combined with vLB correction in the interstitial, the results approach those of optimized mBJ, as well as GW and hybrid functionals.  For example, vLB does not improve the gap for Ge within LMTO-ASA but the FP-NMTO basis set dramatically improves the gap with LDA+vLB, and gets correctly an indirect gap .

{\par}For completeness, we note that we find a large different between results from LDA+vLB (present calculation) and related LB94  (Ref. \onlinecite{pb2018}), which used a linear-augmented plane-wave code ({\tt Wien2K}) with local atomic orbitals. As similar results are found from vLB and  mBJ (which also uses {\tt Wien2K}), we believe that this difference arises from the implementation of vLB asymptotic conditions\cite{Singh2016} and our different basis sets, as discussed in section \ref{Sec:Theory} and section \ref{Sec:Res-Comp} regarding plane-wave results.

\subsection{Application to 2D materials: $h-$BN and $h-$SiC}
The 2D materials, like of graphene and hexagonal boron nitride ($h-$BN), have drawn tremendous attention in terms of both fundamental physics and possible applications in energy-generation devices.\cite{NGPN2009,GN2007} Single layers of graphene and h-BN have been fabricated and found to be stable at room temperature.\cite{NJSBKMG2005,MGKNBR2006,GBBWNG2008} Here, we exemplify the efficacy of our approach by discussing the electronic structure of 2D $h-$BN and SiC.\cite{Jana1,Jana2,Jana3}

\begin{figure}[t]
\includegraphics[scale=0.2]{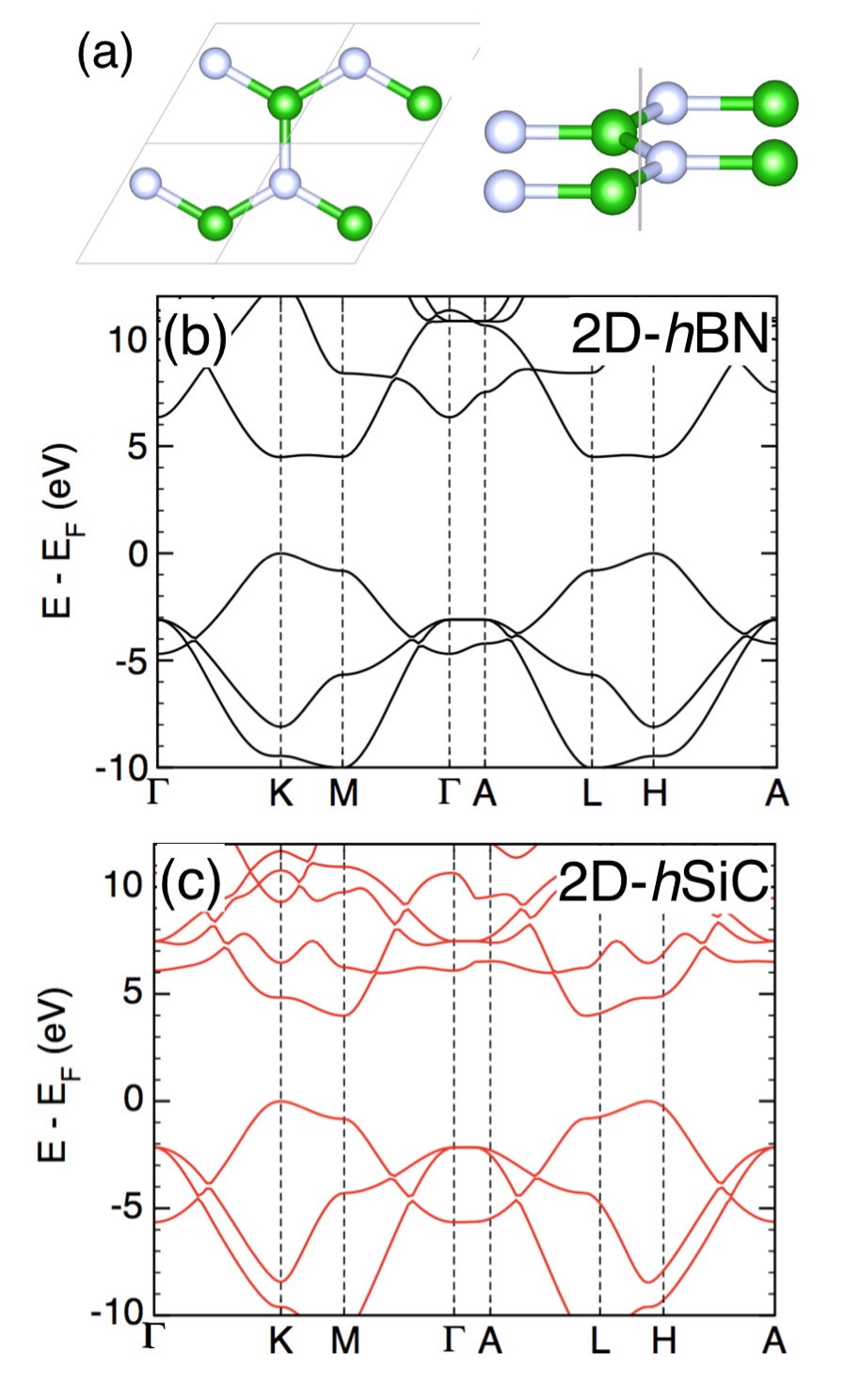}
\caption{(Color online) For 2D $h-$BN and $h-$SiC, (a) (001) and (101) view of crystal structure, and FP-NMTO bands in LDA+vLB  plotted in Brillouin zone along high-symmetry  lines: (b) and (c). Bandgap for $h-$BN ($\sim$4.5 eV) is in good agreement with observed range ($3.6-6$ eV),\cite{hbn} while $h-$SiC (3.90 eV) is overestimated  to observed\cite{lin} (3 eV),  as found in most theories,\cite{azadeh} e.g., G$_0$W$_0$ gets 4.4 eV.\cite{hsueh}}
\label{Fig5}
\end{figure}

{\par}2D $h-$BN is a group III-IV binary compound displaying interesting chemical and electronic structure properties that leads to wide range of technological applications, such as protective coating, deep ultraviolet emitter, and transparent membrane.\cite{Hsu2012,Dai2015} 2D $h-$BN has similar honeycomb atomic structure as graphene with a lattice mismatch of 2\%,\cite{Liu2013,Dean2010} where each primitive cell contains a B and N atom and share total of eight electrons. Each B (N) atom forms $sp^{2}$ hybridized state with N (B), where each state forms three in-plane $\sigma$ (6-electrons) and an out-of-plane $\pi$ (2-electrons) bond. The NMTO-vLB energy minimized lattice constant of 2.508~\AA~is in good agreement with that measured (2.504~\AA).\cite{Solozhenkoa} Unlike graphene,\cite{GN2007} $h-$BN is a wide bandgap insulator.\cite{hbn} The calculated bandgap of $\sim$4.5 eV, in Fig.~\ref{Fig5}(b), falls within the experimentally observed bandgap range of 3.6$-$6 eV.\cite{hbn}  The nature of bandgap can not be clearly stated as the band-energies of the lowest-occupied bands at K and M are the same, i.e., a flat band along (K-M). 

{\par}Similarly, 2D $h-$SiC is a group IV binary compound that can  be viewed as graphene (2D C) or silicene (2D Si) doped with `Si' or `C', respectively. The honeycomb lattice of randomly but homogeneously distributed Si atoms in symmetric, semi-metallic graphene opens up a large  gap. Because of its wide bandgap,  $h-$SiC band structure has been in active study for optoelectronic applications.\cite{BS2017} Unlike the polymorphs of carbon, $h-$SiC is a polar material. In spite of the fact that both constituents of SiC are Group IV elements, charge is transferred from Si to C  due its to higher electronegativity.

{\par}The NMTO-vLB energy minimized lattice constant  of 2.62~\AA~is in good agreement with the measured a=2.6$\pm$0.2~\AA.\cite{azadeh,lin} The bandgap of 3.90 eV (indirect along K to M  in Fig.~\ref{Fig5}(c)) is overestimated compared to experiment (2.96 eV),\cite{lin} a trend is observed in most other theories.\cite{azadeh} We also found good agreement in structural properties with experiments, while other theoretical results show significant differences, which yield a gap to be as high as 4.42 eV due to strong excitonic effects included at the G$_{0}$W$_{0}$ level of theory.\cite{hsueh}

\subsection{ Comparing vLB-corrected Results}\label{Sec:Res-Comp}  
{\par} To avoid confusion when comparing results between different (e.g., site-centered basis vs. plane-wave basis) methods and various implementation of modified semi-local functionals (like vLB) or optimized, screened-range-separated hybrid functionals,  recall that each method has different potential zeroes, $v_o$.  Plane-wave methods use a global reference set to atomic zero ($v_o=0$) far outside the atom or crystal, which requires a dielectric function, $\epsilon$, to solve the macroscopic electrostatics;\cite{PRB2015R}  in particular, to set the asymptotic ($-1/r$) condition in hybrid functional, two range-partition variables must obey a sum rule\cite{PRB2015R}  $(\alpha + \beta)r^{-1}= \epsilon^{-1}r^{-1}$, where $\epsilon=1$ for an atom in vacuum and $1 \le \epsilon \le \infty$ for a crystal. Optimal tuning depends on the values of $\alpha$, $\beta$, and $\epsilon$. Hence, to get the excited states or bandgap, you must calculate $\epsilon$,\cite{PRB2015R} a  significantly more costly calculation.
In contrast, all-electron, site-centered methods can choose to take the potential zero inside the crystal interstitial (eigenvalues relative to $v_o$) and set the potentials asymptotic behavior there. W then only require $I-A$, which is independent of $v_o$. To change the reference to atomic zero to compare, say, to plane-wave results, one requires the work function $W \propto \epsilon$.\cite{Singh2016} Hence, site-centered basis can avoid the calculation of $\epsilon$  for bandgaps still obtain the same result as found from plane-wave calculations.\cite{Singh2016}
For implementations using other methods, like linear-augmented-plane-waves with local atomic orbitals, \cite{mbj,pb2018,mbjp} it is judicious to define the asymptotic value of the potential within the solid.  How the asymptotic behavior is addressed affects results, as found, for example, for LB94, LDA, and LDA+vLB.  Notably,  LDA+vLB should approach the mBJ \cite{mbj}, GW, and screened-hybrid functional \cite{PRB2015R}  results, as we indeed found in Table~\ref{tab1}. Of course, care is needed in comparing results in the solid-state limit.\cite{JCP2015}

\section{Conclusion}
{\par}The bandgaps of IV and III-V semiconductors are underestimated by most semi-local exchange-correlation functionals. To address the problem, we combine (a) self-consistent FP-NMTO method [providing a compact and complete optimal basis throughout a supercell volume -- with no approximations to interstitial regions] with (b) vLB correction to the LDA exchange. The FP-LMTO-vLB approach provides a fast and efficient way to calculate accurate bandgaps, comparable to hybrid exchange functionals, with the speed of semilocal functionals (e.g., GGA). The combined improvement in optimal  basis  set (FP-NMTO) and asymptotic behavior of the exchange hole (LDA+vLB) yield much better bandgaps. The self-consistent FP-NMTO-vLB can be more effective in calculating material properties (e.g., electronic, optical), localized atomic orbitals (Wannier functions), or real-space tight-binding parameters for electronic-structure studies, as exemplified here for bandgaps in group IV and III-V semiconductors and 2D materials.

\section{Acknowledgements}
S.D. and A.M. would like to thank Yoshiro Nohara and O. K. Andersen for permission to modify the FP-NMTO. S.D. is thankful to I. Dasgupta for useful discussion. Research at Ames Laboratory was funded by the U.S. Department of Energy (DOE), Office of Science, Basic Energy Sciences, Materials Science and Engineering Division, which is operated for the U.S. DOE by Iowa State University under contract DE-AC02-07CH11358.

\section*{Appendix:}
\subsection*{Reproducing exact-exchange: The H atom} 
\begin{figure}[h]
\includegraphics[scale=0.16]{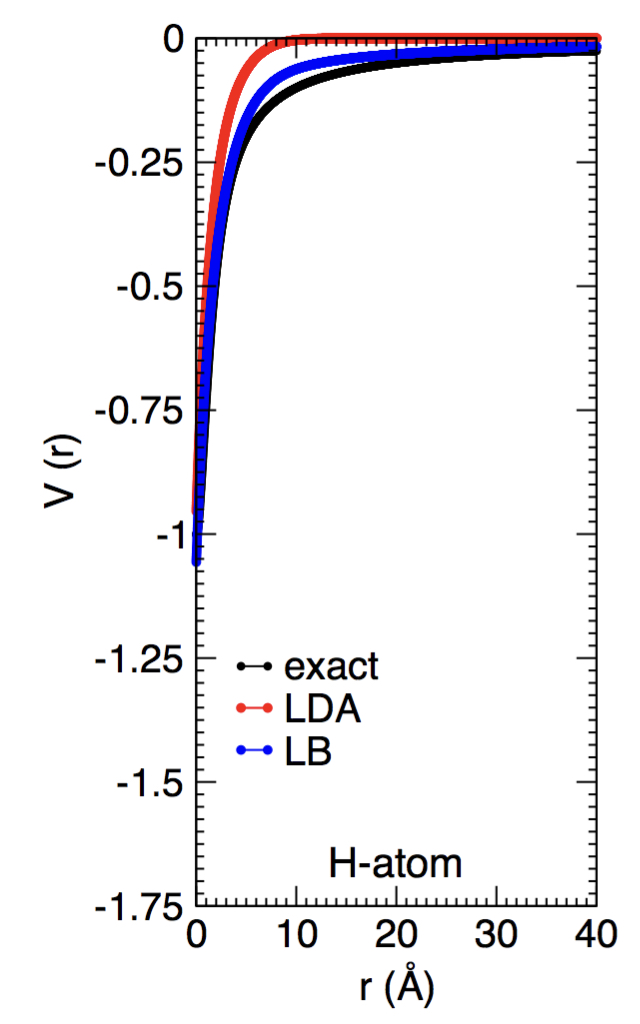}
\caption{(Color online). LDA+vLB compared with the exact-exchange and LDA potentials for a single-electron density.  LDA+vLB matches closely the exact exchange-correlation.}
\label{A1}
\end{figure}

The vLB potential goes as $-1/r$ for densities that decay exponentially as a function of $r$, i.e., as $exp(-r)$, for positions far from the origin. The effective potential V$_{x}+$V$_{c} +$V$_{LB}$ along with the LDA potential V$_{x}+$V$_{c}$ are shown in Fig.~\ref{A1} for an electron in a hydrogen atom. The exact known densities for these systems are used to obtain these potentials. The potentials are compared with the exact XC potential given by negative of the electrostatic potential for these single-electron systems. Except close to $r = 0$, it is clear that the vLB-corrected potential is very close to the exact exchange-correlation potential. 
In  Table~\ref{tab3}, we  display the exact potential (self-interaction free) for a single-electron in a hydrogen (H) atom and compare exact energies calculated with exact, LDA and LB-corrected LDA potentials V$_{exact}$ potentials, where a much smaller error in self-interaction energy of the vLB-corrected potential is notable. Clearly, the LB-corrected LDA potential provides a significant improvement to the energy without invoking need for self-interaction corrections.\cite{Singh2017}

\begin{table}[h]
\begin{tabular}{|c|c|c|c|c|}
\hline
   	 &$V_{exact}$  &   $V_{LB}$   & $V_{LDA}$    \tabularnewline
\hline
	H  	&	0.3144     &	0.2858    &  0.2226     \\
   \%error  	&	    -- 	        &	9.09        &   29.20       \\
\hline 
\end{tabular}
\caption{ {\label{tab3}}  Total energy for 1$e^{-}$ in a H atom calculated using exact, LDA and LDA+vLB potentials. LDA+vLB  yields significant improvement to energy without invoking self-interaction corrections.  Error is given as $\frac{(V_{exact}-V)}{V_{exact}}\times100\%$.} 
\end{table}

 \subsection*{Calculation Details --TB-LMTO-ASA} In TB-LMTO-ASA core states are treated as atomic-like in a frozen-core approximation and energetically higher-lying valence states are addressed in the self-consistent calculations of the effective crystal potential, which is constructed by overlapping Wigner-Seitz spheres for each atom in the unit cell. Two-fold criteria for generating crystal potential, on the same footings of TB-LMTO-ASA, has been used: $\left(a\right)$ use of trial wave function, i.e., linear combinations of basis functions like plane waves in the nearly free-electron method, and $\left(b\right)$ use of matching condition for partial waves at the muffin-tin sphere.\cite{TBLMTO,AJS1986} We have used the LDA correlation parameterized by van Barth and Hedin\cite{LDA} with corrected vLB exchange, matched at the ASA radii. Following atomic-sphere-approximation in TB-LMTO-ASA,\cite{TBLMTO}, the open shell structured semiconductors are filled with empty spheres (ESs) for improved basis. Here, ESs are empty sites with no cores and  small density of electronic charges. The dependence of dimensionless parameter $x$ present in vLB correction on $R_{ASA}$ is very crucial,~so appropriate choice of $R_{ASA}$ is important. In all calculations,~we chose $R_{ASA}$ by $\pm$5-10\% from default values to control the overlapping of atomic spheres and empty spheres to reduce the loss of electrons into the (unrepresented) interstitial for open-shell structures, e.g. semiconductors.

 \subsection*{Timings Comparison for FP-NMTO (vLB) and Quantum Espresso (HSE06)}
{\par} To given some comparison of times, we used FP-NMTO-vLB and Quantum Espresso HSE06 for two cases, Ge and 2D-BN, as shown in Table~{III}.
Even though the methods are quite different (i.e., basis sets, etc.) the hybrid functional is severely costly, in particular a second q-mesh is required for solution.

\begin{table}[h]
\begin{tabular}{|c|c|c|c|c|}
\hline
 {System/Method}  	 & HSE06 (QE)  &   vLB (NMTO)      \tabularnewline \hline
	Ge 	(k-mesh 8x8x8)     &	  250.8     &	6.60         \\ \hline
2D-BN  	(k-mesh 12x12x4) &	    69.6	&	2.76                 \\  \hline 
\end{tabular}
\caption{ {\label{tab3}}  Time (in minutes) for two representative cases: Ge and 2D-BN. We use HSE06 in Quantum Espresso (QE) and LDA+vLB in NMTO on same machine for one-one comparison between two methods.} 
\end{table}

  \subsection*{Wannier Representation in FP-NMTO}
\begin{figure}[h]
\includegraphics[scale=0.13]{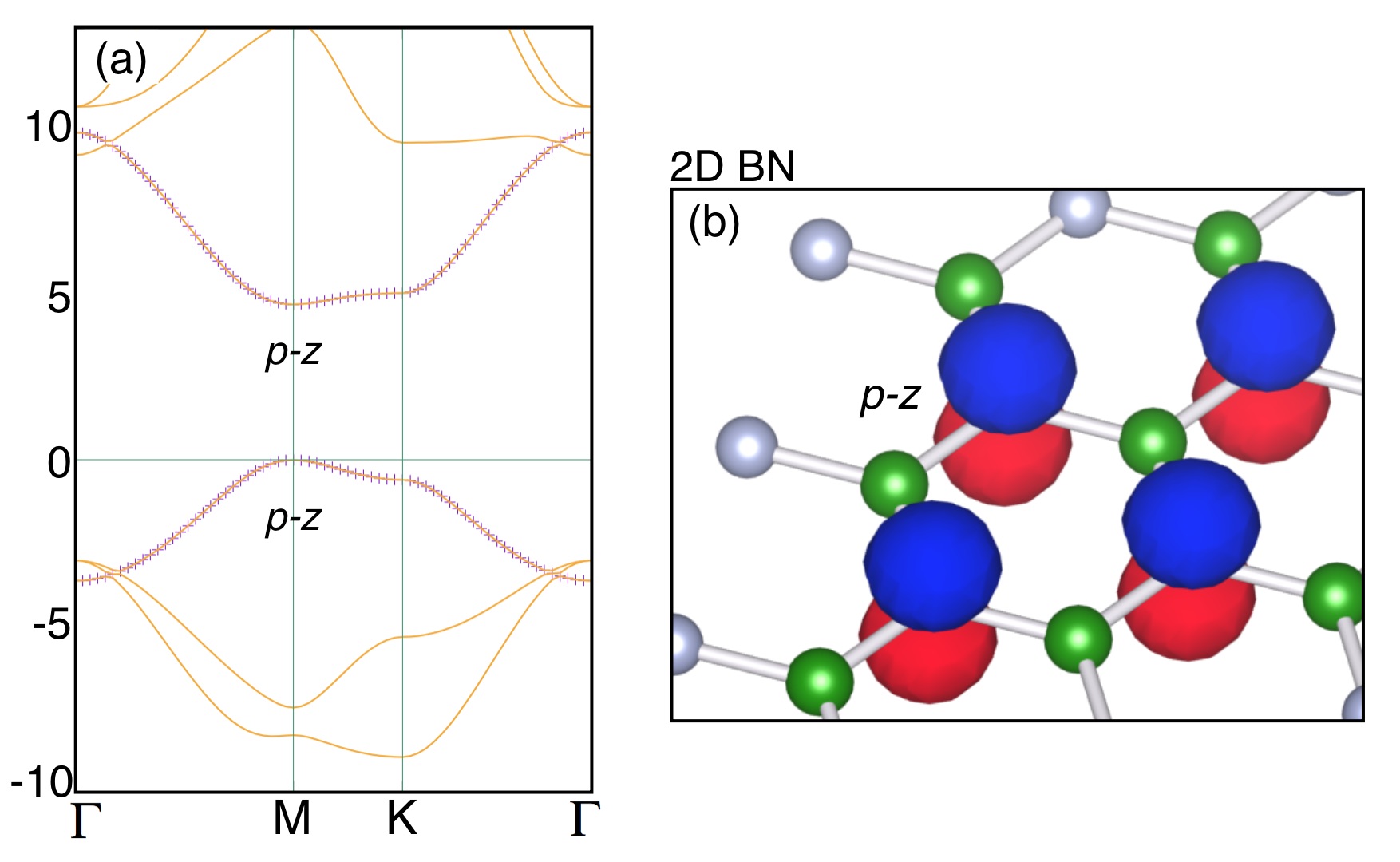}
\caption{(Color online). (a) BN band-structure with down-folded N p$_{z}$ bands (highlighted with red). (b) Wannier function plot of N p$_{z}$ orbitals.}
\label{A2}
\end{figure}

{\par} In Fig.~\ref{A2}, we show a Wannier function representation for the 2D-BN as found within the FP-NMTO using LDA-vLB.

\subsection*{Advantage of FP-NMTO over TB-LMTO-ASA} The improved basis set of FP-NMTO method is accurate, minimal and flexible. Accurate because the FP-NMTO basis solves the KS equation exactly for overlapping muffin-tin potentials. Orthonormalized FP-NMTOs are localized atom-centered Wannier functions, generated in real space with Green-function techniques, without projection from band states.\cite{MWH2012} The TB-LMTO-ASA lacks in all above. Increased computational time is the only downside of the FP-NMTO compared to TB-LMTO-ASA.

\end{document}